\documentclass[a4paper,twocolumn,iop]{emulateapj}
\usepackage{amstext}
\usepackage{apjfonts}
\usepackage{amsmath}
\usepackage{amssymb}
\usepackage{xcolor,xspace}
\usepackage{multirow,threeparttable,tabularx}
\usepackage[colorlinks=true,linkcolor=blue,citecolor=blue]{hyperref}
\usepackage[super]{nth}
\usepackage{ftnxtra}
\usepackage{aas_macros}

\defcitealias{Guillochon:2014a}{GMR14}

\begin{document}

\shorttitle{A Dark Year for Tidal Disruption Events}

\shortauthors{Guillochon and Ramirez-Ruiz}

\title{A Dark Year for Tidal Disruption Events}

\author{James Guillochon\altaffilmark{1,2} and Enrico Ramirez-Ruiz\altaffilmark{3,4}}
\altaffiltext{1}{Harvard-Smithsonian Center for Astrophysics, The Institute for Theory and
Computation, 60 Garden Street, Cambridge, MA 02138, USA}
\altaffiltext{2}{Einstein Fellow}
\altaffiltext{3}{Department of Astronomy and Astrophysics, University of California, Santa Cruz, CA 95064, USA}
\altaffiltext{4}{Radcliffe Fellow}

\email{jguillochon@cfa.harvard.edu}

\begin{abstract} 
Main-sequence disruptions of stars by supermassive black holes result in the production of an extended, geometrically thin debris stream winding repeatedly around the black hole. In the absence of black hole spin, in-plane relativistic precession causes this stream to intersect with itself after a single winding. In this paper we show that relativistic precessions arising from black hole spin can induce deflections out of the original orbital plane that prevent the stream from self-intersecting even after many windings. This naturally leads to a ``dark period'' in which the flare is not observable for some time, persisting for up to a dozen orbital periods of the most bound material, which translates to years for disruptions around black holes with masses $\sim 10^{7} M_{\odot}$. When the stream eventually self-intersects, the distance from the black hole and the angle at which this collision occurs determines the rate of energy dissipation. We find that more-massive black holes ($M_{\rm h} \gtrsim 10^{7} M_{\odot}$) tend to have more violent stream self-intersections, resulting in prompt accretion. For these tidal disruption events (TDEs), the accretion rate onto the black hole should still closely follow the original fallback rate after a fixed delay time $t_{\rm delay}$, \smash{$\dot{M}_{\rm acc}(t + t_{\rm delay}) = \dot{M}_{\rm fb}(t)$}. For lower black hole masses (\smash{$M_{\rm h} \lesssim 10^{6}$}), we find that flares are typically slowed down by about an order of magnitude, resulting in the majority of TDEs being sub-Eddington at peak. This also implies that current searches for TDEs are biased towards prompt flares, with slowed flares likely having been unidentified.
\end{abstract}


\section{Introduction}
The tidal disruptions of main-sequence stars by supermassive black holes are likely responsible for a few dozen flares that have been discovered in the optical \citep[e.g.][]{Velzen:2011a,Gezari:2012a,Chornock:2014a,Arcavi:2014a,Holoien:2014a,Vinko:2015a}, UV \citep{Gezari:2009a}, and X-rays \citep[e.g.][]{Komossa:2004a,Bloom:2011a}. These tidal disruption events (TDEs) show a time-evolution that is consistent with the accretion onto the black hole $\dot{M}_{\rm acc}$ being equal to the fallback rate $\dot{M}_{\rm fb}$ set at the time of the star's disruption (\citealt{Guillochon:2014a}, hereafter \citetalias{Guillochon:2014a}), with the luminosity loosely following a power-law decline of $t^{-5/3}$ \citep{Rees:1988a,Lodato:2009a,Ramirez-Ruiz:2009a,Kesden:2012a,Guillochon:2013a,Cheng:2014a}. However, it remains unclear how these two rates can be so closely related to one another given the difficulty in circularizing material at a distance comparable to the star's periapse distance as dissipation from hydrodynamical effects is limited \citepalias{Guillochon:2014a}.

General relativistic precession can alleviate this issue if the rate of precession is great enough such that the angle of the first stream-stream impact is large, and that the intersection occurs not too far from the black hole such that material is deposited with a semi-major axis comparable to the star's original periapse \citep{Hayasaki:2013a,Bonnerot:2015a,Hayasaki:2015a}. However, the prototypical disruption presented in the literature, a solar-type star being fully disrupted by a $10^{6} M_{\odot}$ black hole, is decidedly un-relativistic, with a periapse distance $r_{\rm p}$ a factor of $\sim 30$ times larger than the gravitational radius $r_{\rm g} \equiv G M_{\rm h} / c^{2}$ for a full disruption \citep{Guillochon:2013a}. Such a disruption does have a stream self-intersection, but at a distance significantly larger than the periapse distance, resulting in a viscous time $t_{\rm visc}$ that can be significantly longer than the timescale of peak fallback accretion $t_{\rm peak}$ (\citealt{Rosswog:2009a}, \citetalias{Guillochon:2014a}, \citealt{Shiokawa:2015a}, \citealt{Bonnerot:2015a}, \citealt{Hayasaki:2015a}).

The effect of black hole spin has been considered for the effect it has on the initial spread in binding energy of the tidal debris \citep{Kesden:2012a}, and for the effect it has on the structure of the debris stream once it has wound around the black hole several times \citep{Stone:2012a,Dai:2013a}. At the moment of disruption, spin can increase (or decrease, depending on if the orbit is prograde or retrograde relative to the black hole's spin) the spread in energy by factors of a few, leading to flares that potentially peak more (or less) quickly than an equivalent Newtonian encounter. In the later phases when the stars have been stretched into a thin stream \citep{Kochanek:1994a}, the structure of the stream, and where it intersects, can be greatly affected by spin. It is this second effect that introduces an additional fixed delay $t_{\rm delay}$ between the time of the star's original periapse and when the tidal disruption is luminous.

In this paper we focus upon determining $t_{\rm visc}$ and $t_{\rm delay}$ for an ensemble of TDEs, and how the values of these parameters both affect the dark period that precedes the start of the flare, and its evolution once accretion begins. In Section \ref{sec:dynamics} we describe the dynamics of the streams resulting from the tidal disruption of a star, and how general relativity (GR) affects these dynamics. In Section \ref{sec:montecarlo} we describe our Monte Carlo approach to constructing a sample of tidal disruption stream configurations in order to determine the statistical properties of their self-interactions. In Section \ref{sec:darkperiod} we present the results of our Monte Carlo exercise, and describe the effect of the stream dynamics upon $t_{\rm delay}$, $t_{\rm visc}$, and the accretion rate $\dot{M}_{\rm peak}$ at time $t_{\rm peak}$. We summarize our results in Section \ref{sec:summary} and describe some caveats of our approach, and what our results imply about the number of identified tidal disruption flares.

\section{Dynamics of Tidal Disruption Streams}\label{sec:dynamics}
Once some material has been liberated from a passing star by the black hole's tides, it is spread in space as a result of the range of binding energies it acquired due to the variance of the tidal force within the star. This spread is equivalent to a spread in velocity of order $\beta v_{\ast}$ relative to the motion of the star's center of mass \citep{Carter:1983a,Stone:2013a}, where $v_{\ast}$ is the star's escape velocity, \smash{$\beta~\equiv~r_{\rm t} / r_{\rm p}$} the star's impact parameter, \smash{$r_{\rm t}~\equiv~q^{1/3} R_{\ast}$} is the tidal radius, $r_{\rm p}$ is the distance of the star at periapse, and $q~\equiv~M_{\rm h}/M_{\ast}$ is the ratio of the black hole's mass $M_{\rm h}$ to the star's mass $M_{\ast}$.

Immediately after the encounter, the self-gravity of the stream continues to play an important role, confining its width to scale as \smash{$r^{1/4}$} (\citealt{Kochanek:1994a}, \citetalias{Guillochon:2014a}), where $r$ is the distance to the black hole. Self-gravity continues to play a role until each part of the stream reaches its apoapse (for the bound portion of the stream); at this point the tidal force applied by the black hole increases relative to the self-gravity term and becomes the dominant component. Because the density of the stream continuously drops as the stream stretches in space, the parts of the stream that have passed their first apoapse are never again confined by self-gravity, and the width of the stream scales homologously, i.e. proportional to $r$. As the internal energy of the stream is so much smaller than its kinetic energy, pressure forces play no role in altering the stream's trajectory, and therefore each part of the stream should closely follow a geodesic within the black hole's spacetime, unless the stream intersects with other parts of itself.

The width of the stream at periapse is determined by the ratio of velocity perpendicular to the orbital plane $v_{\perp}$ to the velocity within the orbital plane $v_{\parallel}$. The first encounter is special; because self-gravity played some role in the stream's evolution, it is somewhat thinner than it would have been otherwise, and thus $v_{\perp}$ is somewhat smaller than $\beta v_{\ast}$. Upon returning to periapse a second time, the stream again experiences compression in which approximately $\beta^{2} \beta_{\rm s}^{2} c_{\rm s,s}^{2}~=~\beta^{2} v_{\ast}^{2}$ in specific kinetic energy is converted into internal energy through a shock, where $\beta_{\rm s}$ is now the effective impact parameter for the stream and $c_{\rm s,s}$ its internal sound speed \citepalias{Guillochon:2014a}. Upon rebound, this energy is converted into kinetic energy in the direction perpendicular to the stream's bulk motion. This process repeats every time a part of the stream returns to periapse.

\begin{figure*}
\centering\includegraphics[width=0.9\linewidth,clip=true]{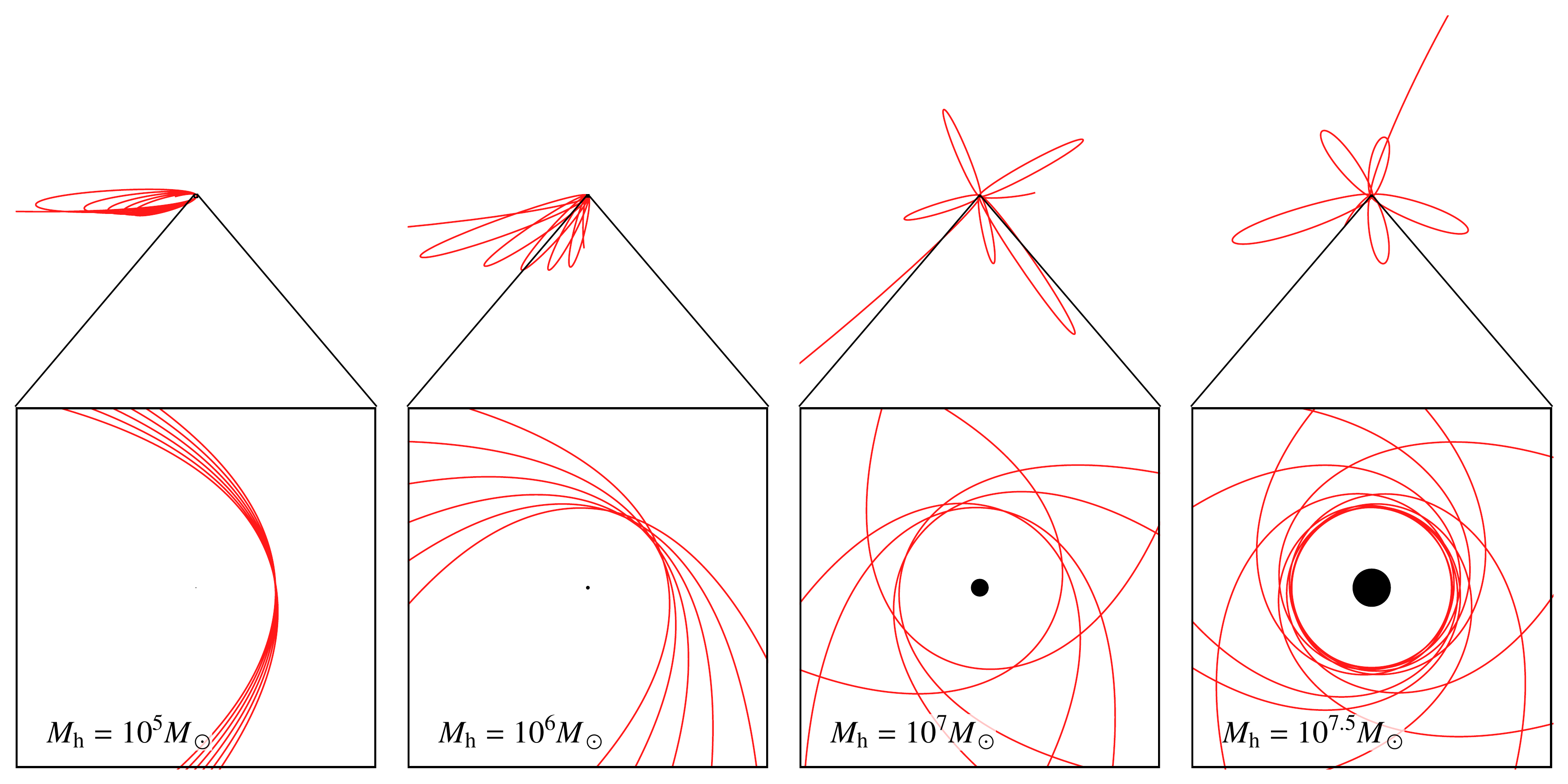}
\caption{Effect of apsidal precession upon the returning debris stream after the disruption of a star by a non-spinning supermassive black hole. The three panels show the structure of the stream after a $\beta = 1$ disruption of a $1~M_{\odot}$ star by black holes of four different masses: $10^{5} M_{\odot}$, $10^{6} M_{\odot}$, $10^{7} M_{\odot}$, and $10^{7.5} M_{\odot}$, with the zoom-in focusing upon a region a factor of four times larger than the star's original periapse. In all panels we show the appearance of the stream a fixed time after the disruption, approximately after the most-bound debris has completed $\sim 5$ orbits, where we have deliberately ignored the effects of stream self-intersection. The black disk in all panels is centered about the black hole and has radius $r_{\rm g}$.}
\label{fig:diagram}
\end{figure*}

While simulations and analytical arguments presented in \citetalias{Guillochon:2014a} suggest that the amount of energy dissipated on the second encounter is comparable to the first encounter, it remains an open question as to how much energy is dissipated on the stream's subsequent encounters. Simulations that have been performed for stars on low-eccentricity orbits \citep{Hayasaki:2013a,Bonnerot:2015a,Hayasaki:2015a} suggest that the width of the stream does not increase dramatically before self-intersection occurs \citep[see Figure 13 of][]{Hayasaki:2013a}. For this paper we presume that the growth in $v_{\perp}$ is linear with the winding number $W$,
\begin{align}
v_{\perp} = W \beta v_{\ast},\label{eq:vperp}
\end{align}
where $\beta v_{\ast}$ is the additional spread in velocity imposed each time the stream returns to periapse. This leads to a stream cylindrical radius of
\begin{equation}
{\cal S} = \frac{r_{\rm p} v_{\rm perp}}{v_{\rm p}} = W \beta r q^{-1/3}\label{eq:streamsize},
\end{equation}
where $W$ is the number of times the stream has returned to periapse. Note that this expression is different from that of \citet{Dai:2013a} who assumed that the stream is self-gravitating even after repeated windings, however this is likely not the case given that the black hole's gravity becomes dominant shortly after the material's first apoapse approach. Equation (\ref{eq:vperp}) should be regarded as a lower limit on the stream width; if anything, the stream may become wider than what we have presumed here given that the shocks formed at the nozzle located at periapse may become increasingly violent as the stream's internal sound speed drops relative to its ballistic speed \citepalias{Guillochon:2014a}. As the details of the stream's width are potentially dependent on cooling and re-ionization \citep{Kochanek:1994a,Hayasaki:2013a,Bonnerot:2015a,Hayasaki:2015a}, the evolution of the stream's width remains an open question that should be addressed by future study.

Because the stream cannot occupy a total solid angle that is greater than 4$\pi$, there is a limit $W_{\rm limit}$ on the number of windings permitted,
\begin{align}
4\pi = \pi \sum_{W = 1}^{W_{\rm limit}} \left[\frac{{\cal S}(r_{\rm p})}{r_{\rm p}}\right]^{2} &= \frac{\pi \beta^{2}}{q^{1/3}} \left[\frac{W_{\rm limit}^{3}}{3} + \frac{W_{\rm limit}^{2}}{2} + \frac{W_{\rm limit}}{6}\right] \\
4\pi &\simeq \frac{\pi \beta^{2} W_{\rm limit}^{3}}{3 q^{1/3}}\\
W_{\rm limit} &= \frac{12^{1/3} q^{1/9}}{\beta^{2/3}}\label{eq:wlimit},
\end{align}
where the second expression assumes that $W_{\rm limit} \gg 1$, although stream self-intersections would likely occur before this limit is reached as the relativistic precession principally occurs within the orbital plane. As we describe below, the first and second windings of the stream almost always intersect in the absence of black hole spin as a result of the apsidal precession (Figure \ref{fig:diagram}, left panel), this is significantly smaller than the limit presented in Equation (\ref{eq:wlimit}).

\subsection{Precession About a Non-Spinning Black Hole}\label{sec:nonspinning}
Ignored thus far have been the effects of GR upon the stream's trajectory. For non-spinning black holes, the amount of precession is dependent upon the ratio of $r_{\rm p}$ to $r_{\rm g}$, and the precession operates prograde to the star's orbit and entirely within its orbital plane, affecting only the argument of periapse $\Omega$. In Figure \ref{fig:diagram} we show the appearance of the stream when $W = 5$ for non-spinning black holes of four different masses, calculated using a freely-available geodesic integrator written for {\tt Mathematica}\footnote{Available at \url{http://goo.gl/FkiUhp}.} \citep{Levin:2008a}. For the lowest-mass case of $M_{\rm h} = 10^{5} M_{\odot}$ (left panel), precession is small, whereas $M_{\rm h} > 10^{7} M_{\odot}$ shows a significant amount of precession, winding more than $2\pi$ radians around the black hole per periapse passage.

Dissipation occurs when the stream strikes itself \citep{Rees:1988a,Kim:1999a,Kochanek:1994a,Ramirez-Ruiz:2009a,Rosswog:2009a,Hayasaki:2013a}, and the amount of dissipation is dependent upon where and at what angle the streams intersect, with the specific energy being dissipated $\Delta E_{\rm s}$ scaling as
\begin{equation}
\Delta E_{\rm s} = \frac{G M_{\rm h}}{r_{\rm int}} \sin^{2} \left[\frac{\alpha_{\rm int}}{2}\right],
\end{equation}
where $r_{\rm int}$ is the distance from the black hole at which the intersection occurs, and $\alpha_{\rm int}$ is the angle at the stream's self-intersection point, which can range from zero to $\pi$. Once this material has struck itself, it acquires a new binding energy to the black hole equal to the amount of energy dissipated (the initial binding energy of the debris is negligible). This sets the semi-major axis of the resulting disk,
\begin{equation}
a_{\rm circ} = \frac{G M_{\rm h}}{2 \Delta E_{\rm s}},
\end{equation}
which has an orbital period of
\begin{equation}
P_{\rm circ} = 2\pi \sqrt{\frac{a_{\rm circ}^{3}}{G M_{\rm h}}}.
\end{equation}
If we assume that this disk is subject to the same magneto-rotational instability that leads to accretion for steadily-accreting black holes, the viscous time (e.g. the time $t_{\rm visc}$ it takes material to accrete) is
\begin{equation}
t_{\rm visc} = \alpha^{-1} \left(\frac{h}{r}\right)^{-2} P_{\rm circ},\label{eq:tcirc}
\end{equation}
where $\alpha$ is the standard viscous parameter and $h$ is the scale-height of the disk \citep{Shakura:1976a}. 

As all material in the stream eventually proceeds through the self-intersection point, all the mass that eventually accretes onto the black hole is subject to the same $t_{\rm visc}$. Because dissipation within the stream is small prior to the stream self-intersection \citepalias{Guillochon:2014a}, the specific binding energy of material to the black hole ${\rm d}m/{\rm d}e$ is equal to that set at the time of disruption. Depending on how $t_{\rm visc}$ compares to the time $t_{\rm peak}$ at which the fallback rate $\dot{M}_{\rm fb}(t)$ would peak, the accretion rate onto the black hole as a function of time $\dot{M}_{\rm acc}(t)$ may or may not closely following $\dot{M}_{\rm fb}(t)$. In general, $\dot{M}_{\rm acc}(t) = \dot{M}_{\rm fb}(t)$ for all times $t > t_{\rm visc}$ \citep{Shiokawa:2015a,Bonnerot:2015a,Hayasaki:2015a}; if $t_{\rm visc} \ll t_{\rm peak}$, the flare's bolometric luminosity should closely follow $\dot{M}_{\rm fb}$ at all times, we label such events as ``prompt.'' If $t_{\rm visc} \simeq t_{\rm peak}$, then the accretion onto the black hole can be slowed at early times, but the fallback at late times and its predicted power-law behaviors \citep{Guillochon:2013a} may be preserved, an event like this is described as being ``rise-affected.'' In cases where $t_{\rm visc} \gg t_{\rm peak}$, the particular shape of $\dot{M}_{\rm acc}$ is unlikely to survive the circularization process, and the flare is prolonged significantly at the expense of reduced peak luminosity, we label such events as ``slowed.'' For slowed flares, the power-law decline is more likely to converge to a universal value of $t^{-1}$ \citep{Cannizzo:1990a}, and thus valuable information regarding the star's original structure set at the time of disruption may be lost.

From Figure \ref{fig:diagram} it is clear that $\alpha_{\rm int}$ for low-mass black holes is small, and first occurs near the material's apoapse, but for higher-mass black holes the amount of precession is so extreme that collisions tend to occur closer to periapse and at larger angles, with multiple points of self-intersection. The only important self-intersection location is the first one, after this point, the stream is widened significantly and diverted from its original path, and the geodesic approximation ceases to be appropriate.

\begin{figure}
\centering\includegraphics[width=\linewidth,clip=true]{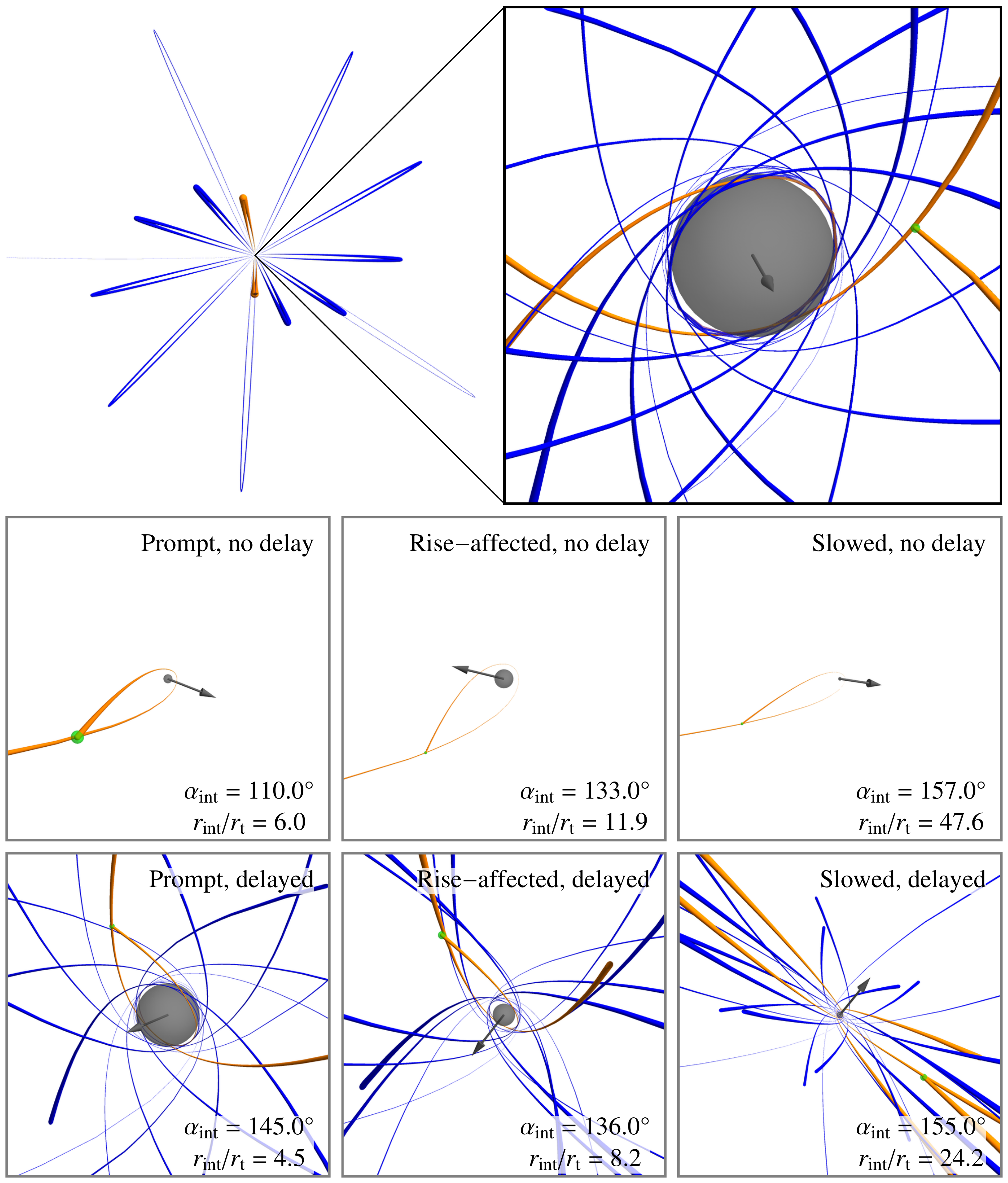}
\caption{Example stream intersections drawn using the Monte Carlo method described in Section \ref{sec:montecarlo} showing the effects of both apsidal and nodal precession. In all panels, the blue ellipses represent parts of the stream that do not intersect any other stream, whereas orange ellipses intersect at least one other ellipse. Intersection points between different parts of the stream are indicated by the small green spheres, whereas the gray spheres show the extent of the innermost bound circular orbit ($r_{\rm IBCO}$). The arrow shows the projection of the direction of the black hole's spin in the image plane, which is rotated to remove the precession given to the star upon its first periapse passage. The top two panels show a wide (left) and close-up (right) view of a stream system in which the most bound material wraps around the black hole 13~times before intersecting itself. The bottom six panels show prototypical examples of the combinations of possible viscous and delay times, with the angle and distance of stream self-intersection being denoted in the bottom right of each panel\footnote{A movie showing all stream configurations in the ensemble is available at \url{http://youtu.be/W3VCdidJ5zg}.}.}
\label{fig:examples}
\end{figure}

\subsection{The Importance of Spin}
When the black hole has spin, the amount of precession now depends on the black hole's spin parameter $a$ and upon the inclination of the orbit relative to the black hole's equatorial plane ${\sc i}$. Additionally, the precession is no longer confined to the orbital plane, resulting in vertical deflection of material. This has an important consequence that is not realized around non-spinning black holes: Whereas any amount of in-plane precession would inevitably result in stream intersection in the non-spinning case (even for a stream of zero width), the nodal precession in a spinning black hole system can deflect material out of its original orbital plane such that an intersection is no longer guaranteed \citep{Dai:2013a}.

The main consequence of this is that there can be a significant delay before stream self-intersection, and thus circularization, can begin. In Figure \ref{fig:examples} we show a small sub-sample of the realizations drawn using the method described in Section \ref{sec:montecarlo} to demonstrate how the nodal precession causes the stream to miss itself, with a collision sometimes occurring after many windings (Fourteen in the particular example presented in the top panel). But whereas a viscous time in excess of the fallback time can alter the shape of the resulting accretion function $\dot{M}_{\rm acc}(t)$, the delay introduced by the stream avoiding self-intersection does {\it not} affect the shape of $\dot{M}_{\rm acc}(t)$. This is because self-gravity, internal pressure, and dissipation at the nozzle-point are all negligible as compared to the binding energy of the stream to the black hole \citep{Luminet:1985a}, and none of these effects introduce appreciable corrections to $\dot{M}_{\rm acc}(t)$ even if the stream stretches by factors of tens of thousands relative to the star's original size.

In combination with the ``prompt,'' ``rise-affected,'' and ``slowed'' flares described in Section \ref{sec:nonspinning}, the introduction of a potentially significant delay caused by the prolonged period of time before the stream self-intersects yields six possible circularization behaviors, summarized in the bottom six panels of Figure \ref{fig:examples}. A flare may be for instance ``prompt'' in the sense that its viscous time is short as compared to $t_{\rm peak}$, but circularization may not begin for years after the star was first disrupted.

Estimating the amount of delay per system is complicated, as it depends not only on the physical parameters of the black hole and the star (mass, spin, size), but also upon the orientation of the star's orbit, and on how the stream's size grows with each periapse passage (Equation (\ref{eq:streamsize})). A system is likely to have a delay if the precession rate is large enough to cause the stream to miss itself at least once; this implies that the nodal precession per orbit $\Delta \omega$ must exceed an angle greater than the angular size of the accretion stream upon its return to periapse \citep{Stone:2012a},
\begin{equation}
\Delta \omega > \frac{R_{\ast}}{r_{\rm p}} = \frac{\beta q^{-1/3}}{2 \pi}.
\end{equation}
We perform a Monte Carlo calculation to determine how often this condition is satisfied, and what effects this deflection has upon TDE observables.

\section{Monte Carlo Realizations of Tidal Disruption Streams}\label{sec:montecarlo}
The trajectories presented in Figure \ref{fig:diagram} were generated integrating the full geodesic equations around a non-spinning black hole, and the code that we utilized is also capable of integrating geodesics about spinning black holes. However, generation of these curves is computationally expensive given that every element of the tidal disruption stream has a different binding energy and angular momentum, requiring that all points in the stream need to be integrated separately \citep[alternative methods that avoid direct integration may be faster, see e.g.][]{Dexter:2009a}. Additionally, it is difficult to derive a robust algorithm for finding when the stream self-intersects; the distance between every point on the curve needs to be evaluated between every other point, meaning that the computational cost scales with the square of the number of points used to resolve the stream. This is further complicated by the fact that the stream width scales both with $r$ and $W$, the determination of which requires tracking the number of times each fluid parcel has returned to periapse.

\begin{figure*}
\centering\includegraphics[width=0.85\linewidth,clip=true]{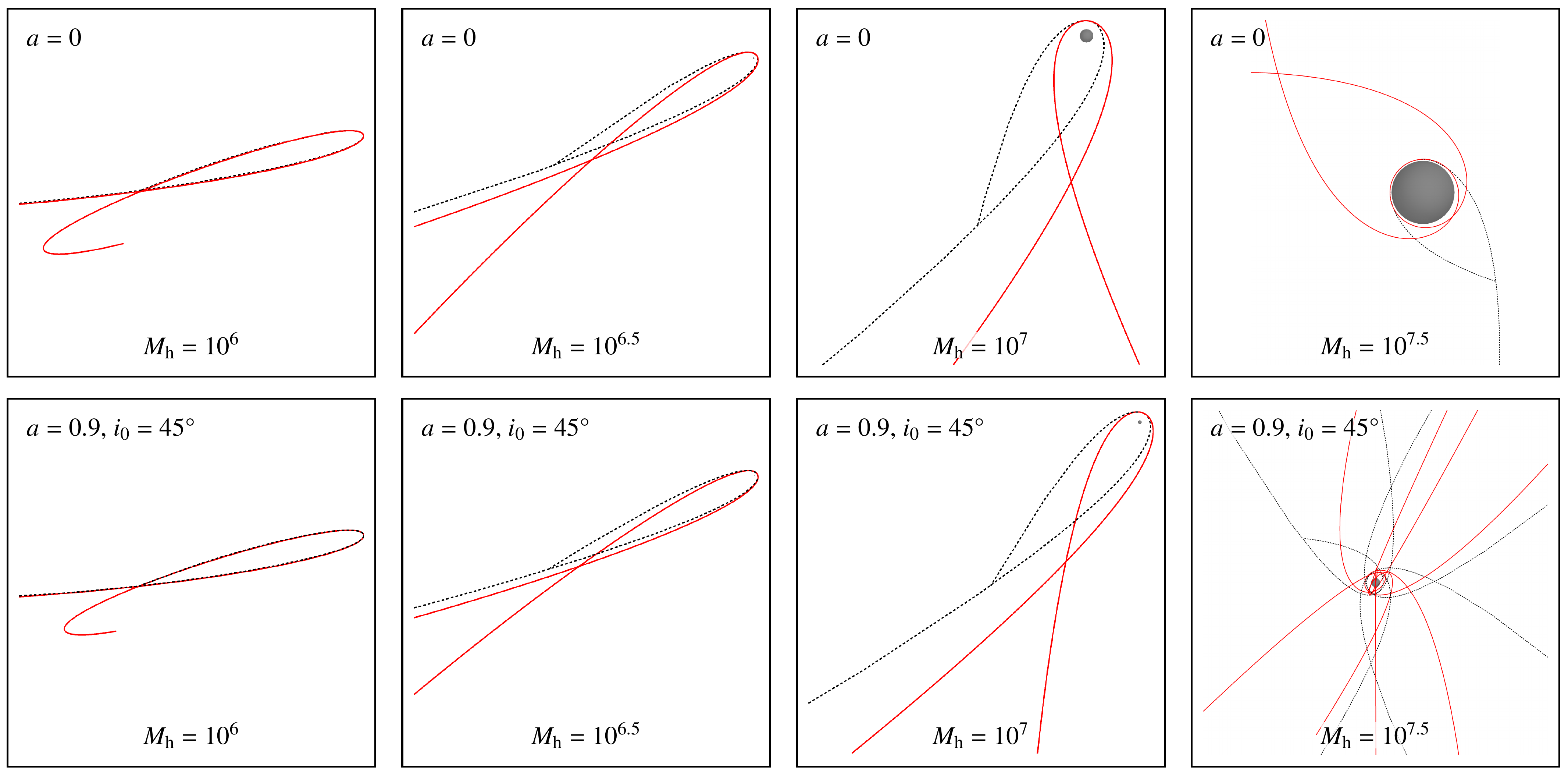}
\caption{Comparison between heuristic prescription (as described in Section \ref{sec:montecarlo}) and exact solutions for non-spinning (top row) and spinning (bottom row) black holes of different masses, where $i_{0}$ is the star's initial inclination relative to the black hole spin plane. Each panel shows two-dimensional projections of the exact solution as the dashed black curve, and the heuristic solution with the solid red curve and, with the heuristic curve terminating once a self-intersection has occurred, and the exact curve being evaluated such that the number of windings is equal to the heuristic solution. The gray sphere in each panel shows $r_{\rm IBCO}$. For clarity, the line thickness in this Figure is not equal to the stream's true thickness.}
\label{fig:compare}
\end{figure*}

As we are interested in generating a sample that's large enough to perform statistics, we want to avoid having to solve many differential equations and performing an expensive calculation for each realization. Every fluid element in the tidal disruption stream possesses a very large eccentricity $\gtrsim 1 - q^{-1/3}$, as most tidally-disrupted stars originate from the sphere of influence where their initial binding energy to the black hole is close to zero \citep{Magorrian:1999a,Wang:2004a,MacLeod:2012a}. As a consequence, practically all of the precession resulting from relativistic effects occurs within a few periapse distances, with the trajectory following a Keplerian ellipse beyond these regions (Figure \ref{fig:diagram}). This means that the general shape of the resulting stream is well-approximated by a series of ellipses where each ellipse has precessed by some amount relative to ellipses corresponding to material with longer periods.

This motivates our method where we represent the tidal disruption stream heuristically by a series of unconnected ellipses. Such configurations are trivial to generate rapidly, only requiring the inclination ${\sc i}$ and eccentricity $e$ of each ellipse to be substituted into the post-Newtonian expressions for precession \citep{Merritt:2013a}

\begin{align}
\Delta \omega &= \left(\Delta \omega \right)_{\rm J} + \left(\Delta \omega \right)_{\cal Q}\label{eq:nodprec}\\
\Delta \Omega &= \left(\Delta \Omega \right)_{\rm D} + \left(\Delta \Omega \right)_{\rm J} +\left(\Delta \Omega \right)_{\cal Q}\label{eq:apsprec}\\
\left(\Delta \omega \right)_{\rm J} &= \frac{4\pi a}{c^{3} q^{1/2}} \left[\frac{G M_{\rm h} \beta}{R_{\ast} (1+e)}\right]^{3/2}\\
\left(\Delta \omega \right)_{\cal Q} &= \frac{3\pi a^{2}}{c^{4} q^{2/3}} \left[\frac{G M_{\rm h} \beta}{R_{\ast} (1+e)}\right]^{2} \cos {\sc i}\\
\left(\Delta \Omega \right)_{\rm D} &= \frac{6\pi G M_{\rm h} \beta}{c^{2} q^{1/3} R_{\ast} (1+e)}\\
\left(\Delta \Omega \right)_{\rm J} &= -4 \cos {\sc i} \left(\Delta \omega \right)_{\rm J}\\
\left(\Delta \Omega \right)_{\cal Q} &= \frac{1 - 5 \cos^{2} {\sc i}}{2 \cos {\sc i}} \left(\Delta \omega \right)_{\cal Q},
\end{align}
where $\Delta \omega$ and $\Delta \Omega$ are the orbit-averaged nodal and apsidal precession terms, and where we have included the de Sitter terms (subscript D), Lense-Thirring terms (subscript J), and quadrupole terms (subscript ${\cal Q}$).

There are two qualitative behaviors that are lost by this approach. Firstly, the ellipses do not capture the ``zoom-whirl'' behavior, which manifests when $r_{\rm p}$ is only a factor of a few larger than $r_{\rm g}$; an example zoom-whirl is visible in the right-most panel of Figure \ref{fig:diagram}. However, when spin is introduced, the probability of self-intersection within the zoom-whirl region is small as most of the nodal deflection is occurring there. Secondly, because the ellipses are unconnected, we do not explicitly resolve the small connecting segments between neighboring ellipses. These segments are only $\sim r_{\rm p}$ long and thus represent a small fraction of the stream's total length, and they are somewhat accounted for by the periapse portions of each ellipse. Over a large ensemble of realizations, we would expect that this small difference would not affect the probability that a self-intersection occurs for a given set of physical parameters.

The system of ellipses is constructed as follows: For each orbit of the most-bound material, a new ellipse is created with a semi-major axis equal to that of the most-bound material $a_{\rm mb}$. Ellipses that were previously created are rotated according to the precession terms defined in Equations~(\ref{eq:nodprec})~and~(\ref{eq:apsprec}), with their semi-major axes being set to $a_{\rm mb} (W_{\rm mb} - W + 1)^{2/3}$, where $W_{\rm mb}$ is the number of windings of the most-bound material.

The orbit-averaged expressions given above are only accurate to $O(v^{2}/c^{2})$ and diverge from the true solutions for very relativistic encounters. As an example, the expression that is expanded via a Taylor series to produce the De Sitter term $(\Delta \Omega)_{\rm D}$ becomes imaginary for $r < 6 r_{\rm g}$ and is maximally $2 \pi$, in reality the precession angle can be arbitrarily large \citep{Levin:2009a}. Qualitatively, we find that the precession terms slightly underestimate the degree of precession when in the weak relativistic regime (leftmost panels of Figure \ref{fig:compare}). Some additional error is also introduced by the orbit-averaged rotations being performed in some order. We choose to perform the apsidal rotation followed by the nodal rotation, however the apsidal and nodal rotations are in reality occurring at the same time. Both of these simplifications introduce some error in the quantities that are expected to be conserved: The component of the angular momentum $L_{z}$ parallel to the black hole's spin axis, and the Carter constant $Q$ \citep{Drasco:2004a}. We find that the error in $L_{z}$ and $Q$ is $< 10\%$ for 93\% of realizations, with the greatest deviations occurring for systems in which $r \lesssim 6 r_{\rm g}$. Because the orbital energy for each ellipse is set by the number of windings, which depends on the time since disruption, orbital energy is perfectly conserved by construction.

Figure \ref{fig:compare} shows a comparison between our heuristic approach and the exact solution for an array of orbits about spinning and non-spinning black holes. From this figure the heuristic solution can be seen to be fairly close to the exact solution for $M_{\rm h} \lesssim 10^{\smash{6.5}} M_{\odot}$, with significant deviations beginning to arise for more massive black holes. However, despite the differences in orbital shape, the location and angle of the first stream self-intersection is very similar for the exact and heuristic approaches, with the exact solutions tending to produce self-intersections that occur slightly closer to the black hole and at slightly larger angles. The biggest qualitative disagreement comes for very massive, non-spinning black holes (upper right panel of Figure \ref{fig:compare}) in which the orbit undergoes a zoom-whirl behavior, resulting in the stream striking itself very close to the event horizon. However, as we find below, most black holes in which such extreme in-plane precession exists are likely to be spinning fast enough to deflect the stream out of its original orbital plane.

With the creation of each ellipse, a root-find is performed between all pairs of ellipses (labeled $i$ and $j$) to find the minimum of the following expression,
\begin{equation}
{\cal D}({\bf x}_{i},{\bf x}_{j}) = \left|{\bf x}_{i}-{\bf x}_{j}\right| - {\cal S}(\left|{\bf x}_{i}\right|) - {\cal S}(\left|{\bf x}_{j}\right|),
\end{equation}
where ${\cal S}$ is calculated using Equation (\ref{eq:streamsize}). If ${\cal D} < 0$, the two ellipses are considered to be overlapping and a stream self-intersection has occurred. In the event that two ellipses intersect at multiple locations or that more than two ellipses intersect (e.g. the \nth{2} winding intersects with both the \nth{4} and \nth{5} windings), we only consider the intersection that occurs chronologically first, as presumably the stream would be deflected significantly at the location of this first intersection. The viscous time $t_{\rm visc}$ is then calculated using Equation \ref{eq:tcirc}, with $r_{\rm int}$ and $\alpha_{\rm int}$ being determined at the point where the two ellipses overlap. The delay time is calculated by the number of windings prior to the first self-intersection,
\begin{equation}
t_{\rm delay} = W P_{\rm mb},
\end{equation}
where $P_{\rm mb}$ is the orbital period of the most-bound material.

We draw random parameters for 4,096 systems assuming a flat distribution in $\log_{10} M_{\rm h}$ with $5\,<\,\log_{10} M_{\rm h}\,<\,8$, a Kroupa distribution for $M_{\ast}$ \citep{Kroupa:1993a} with $-1~<~\log_{10} M_{\ast}\,<\,2$, a flat distribution in black hole spin $a$, an impact parameter distribution that scales as $\beta^{-2}$ \citep{Rees:1988a}, and random initial inclinations ${\sc i}$ relative to the black hole's spin plane. Stellar radii are calculated using the zero-metallicity main-sequence mass-radius fitting formula of \citet{Tout:1996a}. We check that the periapse of each star lies outside the black hole's IBCO \citep{Bardeen:1972a}, if not we redraw $\beta$ and ${\sc i}$, keeping other parameters fixed, until the periapse lies outside this region.

\begin{figure}
\centering\includegraphics[width=0.8\linewidth,clip=true]{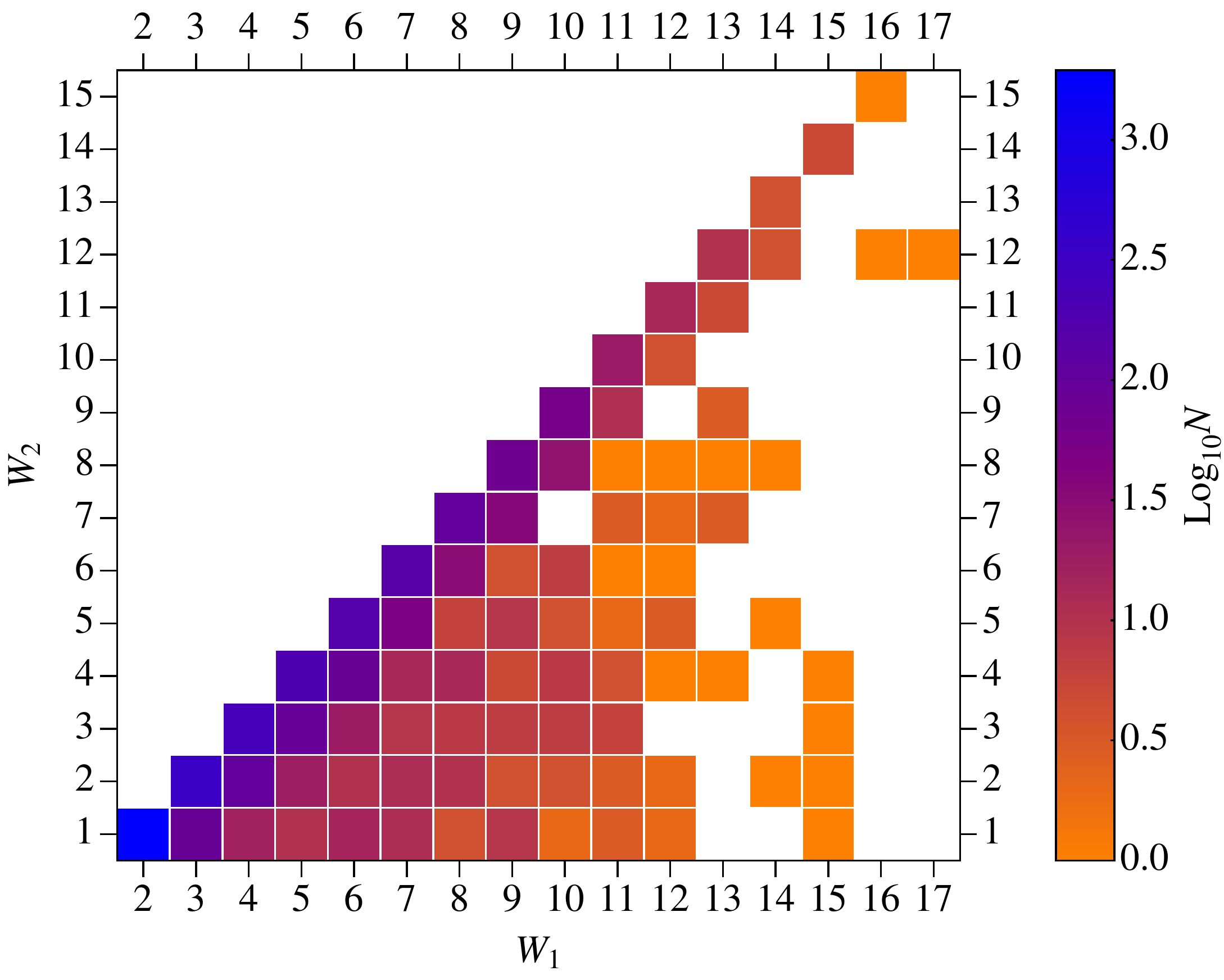}
\caption{Density histogram of realizations as a function of the winding numbers of the two parts of the stream that intersect one another, where $W_{1}$ and $W_{2}$ refer to the winding number of each ellipse. As a single ellipse cannot intersect itself, $W_{1} = W_{2}$ is not permitted.}
\label{fig:windings}
\end{figure}

\section{Delays and Slowdowns for Tidal Disruption Events}\label{sec:darkperiod}
Our Monte Carlo exercise allows us to calculate the duration of the dark period $t_{\rm delay}$ and the viscous time $t_{\rm visc}$ for a random ensemble of TDEs. We find in general that flares with a significant $t_{\rm delay}$ tend to have their first stream self-intersection once the most-bound material has wrapped several times around the black hole, with a half-dozen windings being common, and seventeen windings being the maximum found in our ensemble of realizations (Figure \ref{fig:windings}). Usually the stream self-intersection occurs between neighboring windings, e.g. the \nth{5} winding is likely to have an intersection with the \nth{6} winding, but unlikely to with the \nth{12} winding. This implies that that ratio of densities between the two windings are likely to be similar to one another, and thus these self-intersections should lead to significant dissipation. Events with the smallest $t_{\rm delay}$ values tend to be those in which the \nth{1} and \nth{2} windings intersect, in other words these events begin circularizing once the most bound material completes a single orbit around the black hole.

\begin{figure}
\centering\includegraphics[width=\linewidth,clip=true]{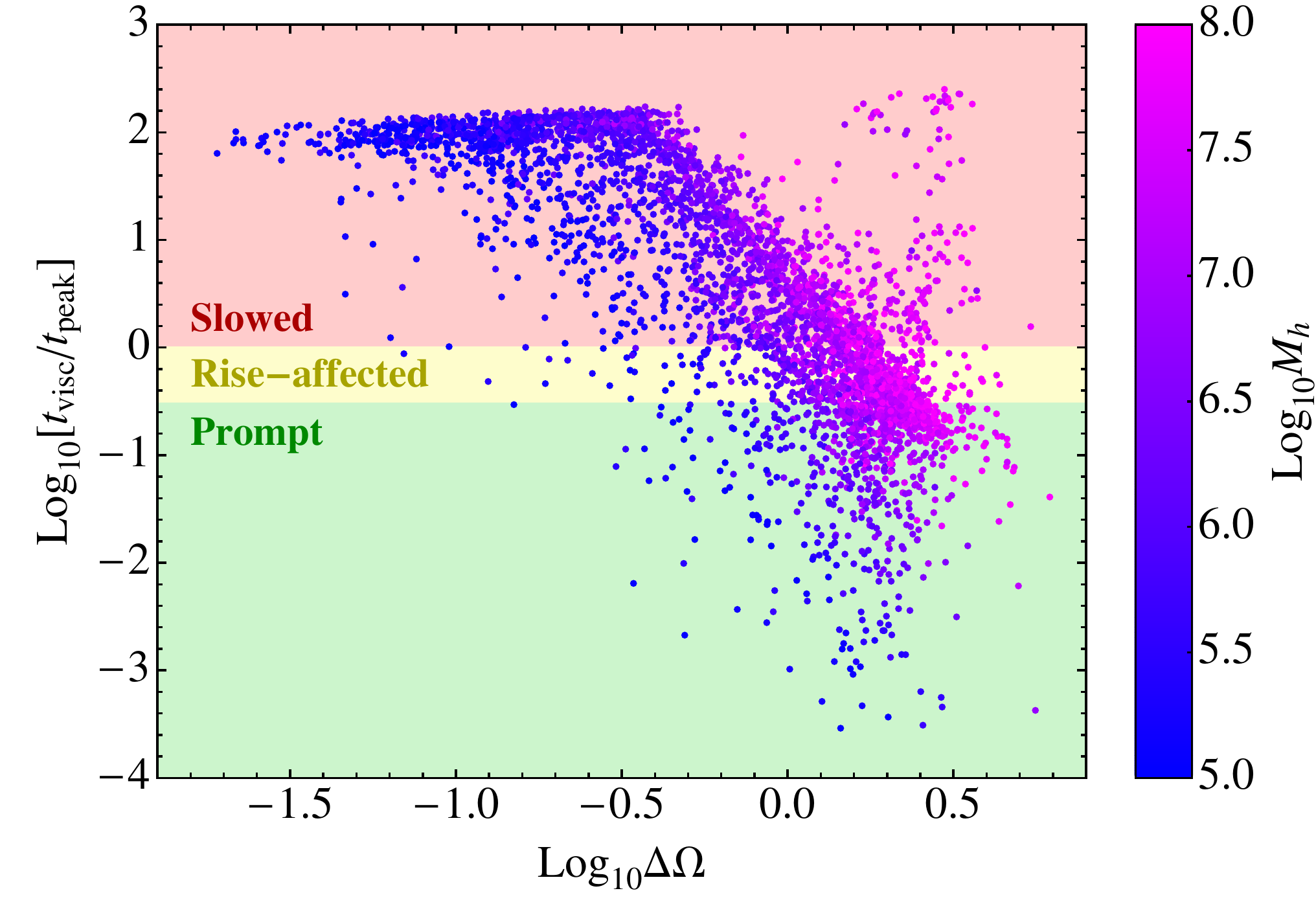}
\caption{Dependence of $t_{\rm visc}/t_{\rm peak}$ on the amount of apsidal ($\Delta \Omega$) precession per orbit. The background coloring corresponds to the circularization behavior described in Section \ref{sec:darkperiod}, with the color of the points corresponding to $\log_{10} M_{\rm h}$ as indicated by the colorbar.}
\label{fig:visc}
\end{figure}

In Figure \ref{fig:visc} we show how $t_{\rm visc}$ and $t_{\rm delay}$ relates to the time of peak accretion $t_{\rm peak}$, where we use fitting relations \citep{Guillochon:2013a,Guillochon:2015a} to determine $t_{\rm peak}$. For the purposes of calculating $t_{\rm visc}$, we need to assume the parameters that govern the viscous evolution of the accretion disk once it has formed. For the figures and fitting formulae in this paper we select values appropriate to a thick disk, $\alpha = 0.1$ and $(h/r)^{2} = 0.1$, yielding a viscous time that is 100 times longer than the orbital period (this assumption is relaxed in Section \ref{sec:summary}). We consider an event to be ``prompt'' (green color in the figures) if $3 t_{\rm visc} < t_{\rm peak}$, to be ``rise-affected'' (yellow) if $t_{\rm visc} < t_{\rm peak} < 3 t_{\rm visc}$, and to be ``slowed'' if $t_{\rm peak} < t_{\rm visc}$.

\begin{figure}
\centering\includegraphics[width=\linewidth,clip=true]{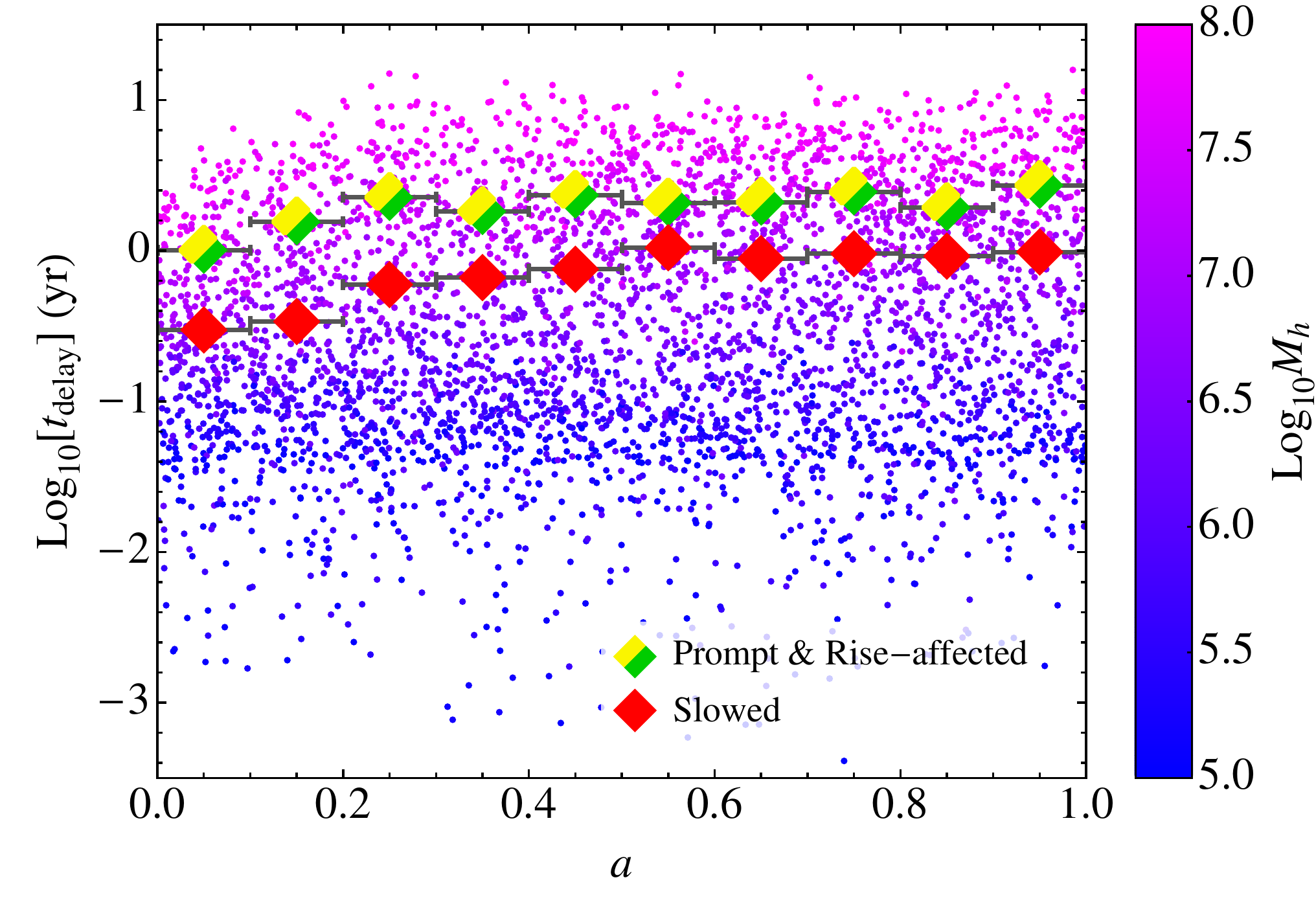}
\caption{Dependence of $t_{\rm delay}$ upon the black hole's spin parameter $a$. Each point shows an individual realization, with the color of the points corresponding to $M_{\rm h}$ as indicated by the color bar. The colored diamonds show the mean values within a range of spin parameter indicated by their error bars, where we have combined the prompt and rise-affected categories into the yellow/green diamonds, and the slowed events are shown by the red diamonds.}
\label{fig:delay}
\end{figure}

Figure \ref{fig:visc} shows that $t_{\rm visc}/t_{\rm peak}$ is closely related to the amount of apsidal precession per orbit, and because this generally increases with increasing \smash{$M_{\rm h}$ as $r_{\rm p}/r_{\rm g}\,\propto\,M_{\rm h}^{-2/3}$}, the viscous times tend to be shorter for more-massive black holes. There are exceptions even for large values of $\Delta \Omega$ which still have large $t_{\rm visc}$ values, these are frequently systems with long delay times that happen to have stream self-intersections with small intersection angles $\alpha_{\rm int}$ (e.g. lower right panel in Figure~\ref{fig:examples}). Because some stream self-intersections occur at nearly zero $\alpha_{\rm int}$ resulting in little dissipation, $t_{\rm visc}$ is set to be the minimum between Equation (\ref{eq:tcirc}) and the time required for the combined cross section of all parts of the stream to cover the full $4\pi$ steradians,
\begin{equation}
t_{\rm cover} = P_{\rm mb} W_{\rm limit},
\end{equation}
where $W_{\rm limit}$ is calculated using Equation (\ref{eq:wlimit}) and $P_{\rm mb}$ is the period of the most-bound material. This typically caps $t_{\rm visc}$ to be no more than a factor of 100 times greater than $t_{\rm peak}$.

The time of delay $t_{\rm delay}$ between when the disruption of the star occurs and the accretion of matter onto the black hole is shown in Figure \ref{fig:delay} as a function of the spin parameter $a$. While there is a lot of scatter at a given $a$, as one might expect given the range of initial inclinations relative to the spin plane, there is a clear reduction in mean values of $t_{\rm delay}$ for events with $a \lesssim 0.2$. This suggests that there is a minimum $a$ required to deflect the stream enough to avoid striking itself in the first few windings, but once above this threshold, the spin of the black hole does not have a strong influence on $t_{\rm delay}$. Because the observed spins of supermassive black holes tend to be greater than 0.5 \citep{Reynolds:2014a} as opposed to our assumption of a flat distribution of $a$, this suggests that most black holes possess enough spin to induce the deflections necessary to cause a delay. As $a < 0.2$ black holes constitute only 20\% of our sample, we find that their exclusion does not alter the statistical properties of the ensemble, and thus our results should be directly applicable to the observed, higher-spin sample. We also find that $t_{\rm visc}$ is very slightly shorter for larger $a$, suggesting that the main effect of $a$ is to increase the amount of time between disruption and circularization, but has little effect on the circularization duration.

\begin{figure}
\centering\includegraphics[width=\linewidth,clip=true]{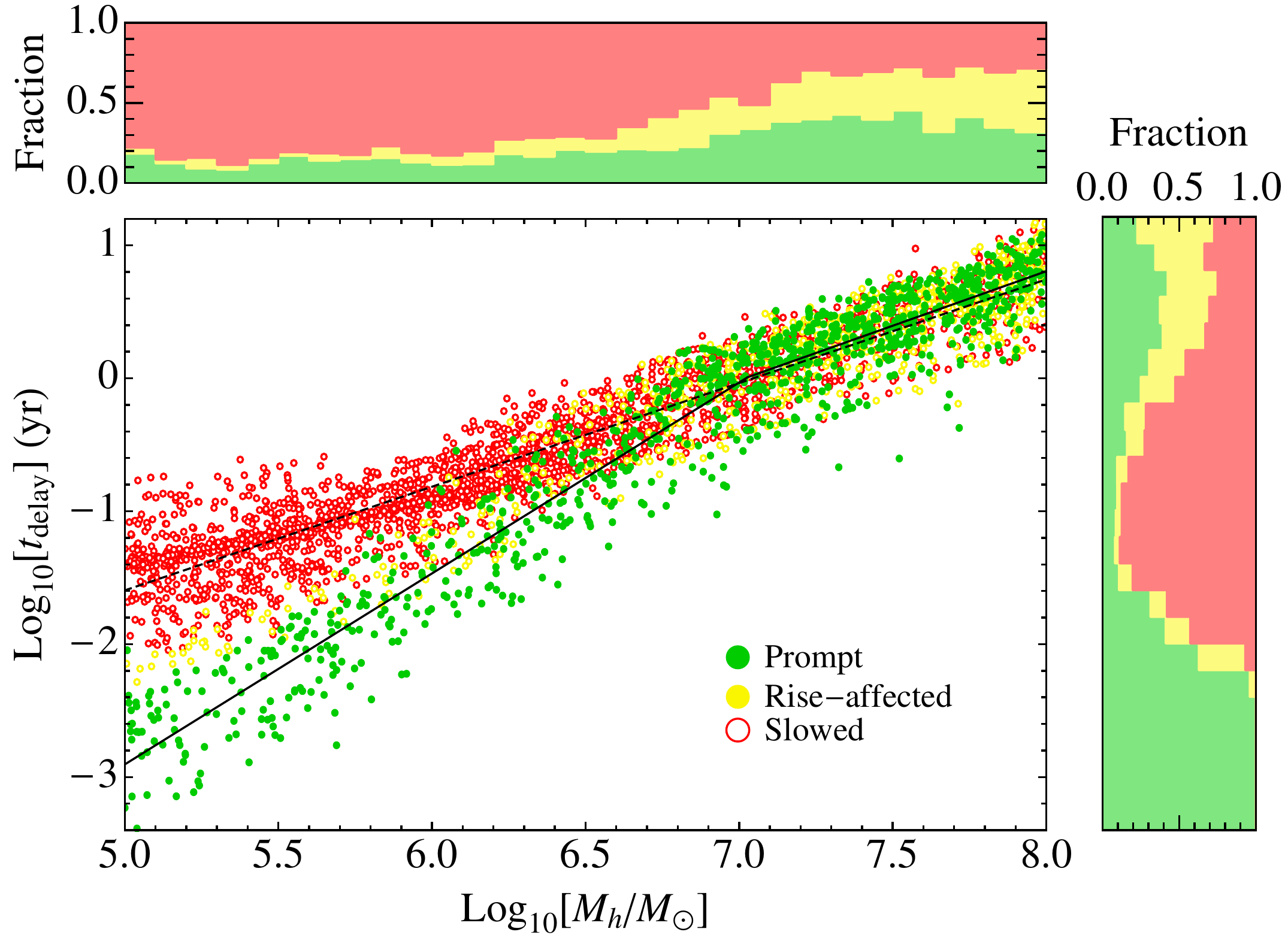}
\caption{Length of the delay period $t_{\rm delay}$ as a function of black hole mass $M_{\rm h}$. The points are color-coded according to how the viscous time $t_{\rm visc}$ compares to $t_{\rm peak}$, the time of peak for the fallback rate. Prompt events are represented by the filled green points, events with an affected rise by the filled yellow points, and slowed events by the open red points, these classes are defined in Section \ref{sec:darkperiod}. The histograms on the right and top sides of the figure show the fraction of events belonging to each category. The solid lines show a least-squares fit to only the prompt flares, whereas the dashed line shows a fit to the slowed events only.}
\label{fig:tdelay}
\end{figure}

In Figure \ref{fig:tdelay} we show a scatter plot of $t_{\rm delay}$ as a function of $M_{\rm h}$, where the color-coding corresponding to its circularization behavior. With our assumptions on the viscous time, we find that there is a change in the fraction of prompt events at a black hole mass of $10^{7.0} M_{\odot}$, with the majority of events being slowed below this mass and the majority being prompt or rise-affected above it. If the viscous time were to be shorter than what we have assumed, the transition mass would move to lower black holes masses and more events will be considered prompt, whereas a longer viscous time will slow down prompt events occurring around black holes of higher mass.

We perform a least-squares fits to $t_{\rm delay}$ to determine the trends in $t_{\rm delay}$ as a function of $M_{\rm h}$. For events that are slowed, we find that the delay time scales sub-linearly with the black hole's mass,
\begin{equation}
t_{\rm delay,slow} = 0.92 \left(\frac{M_{h}}{10^{7} M_{\odot}}\right)^{0.78}~{\rm yr},\label{eq:tdelayslow}
\end{equation}
whereas we find that for events that are prompt follow a broken power-law behavior, with events falling below the cutoff black hole mass having $t_{\rm delay}$ scaling super-linearly with $M_{\rm h}$,
\begin{equation}
t_{\rm delay,fast} =
\begin{cases}
0.93 \left(\frac{M_{h}}{10^{7} M_{\odot}}\right)^{1.44}~{\rm yr},&M_{\rm h} < 10^{7.0} M_{\odot}\\
0.97 \left(\frac{M_{h}}{10^{7} M_{\odot}}\right)^{0.82}~{\rm yr},&M_{\rm h} > 10^{7.0} M_{\odot}
\end{cases}.\label{eq:tdelayfast}
\end{equation}

\begin{figure*}
\centering\includegraphics[width=0.45\linewidth,clip=true]{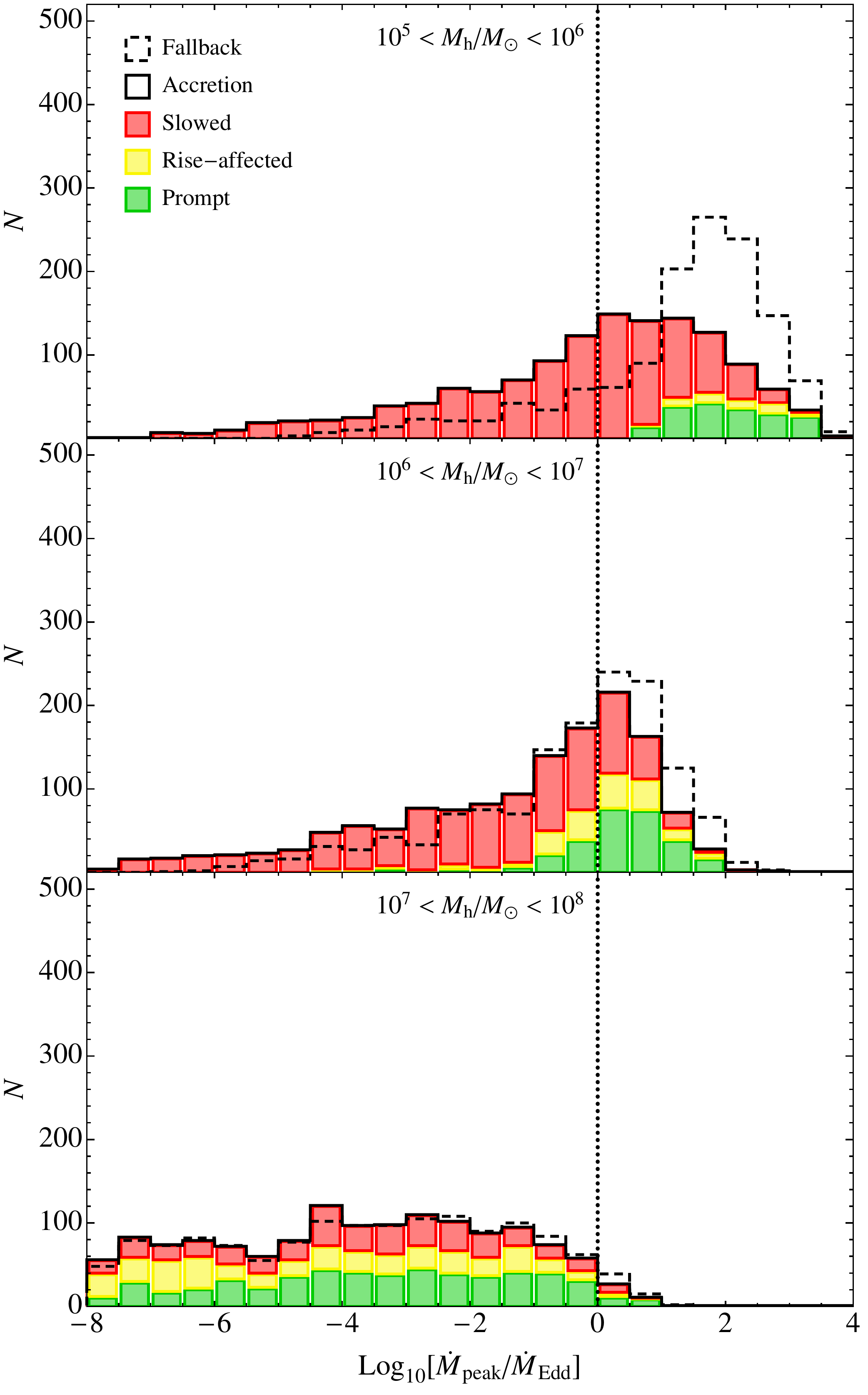}\quad\includegraphics[width=0.45\linewidth,clip=true]{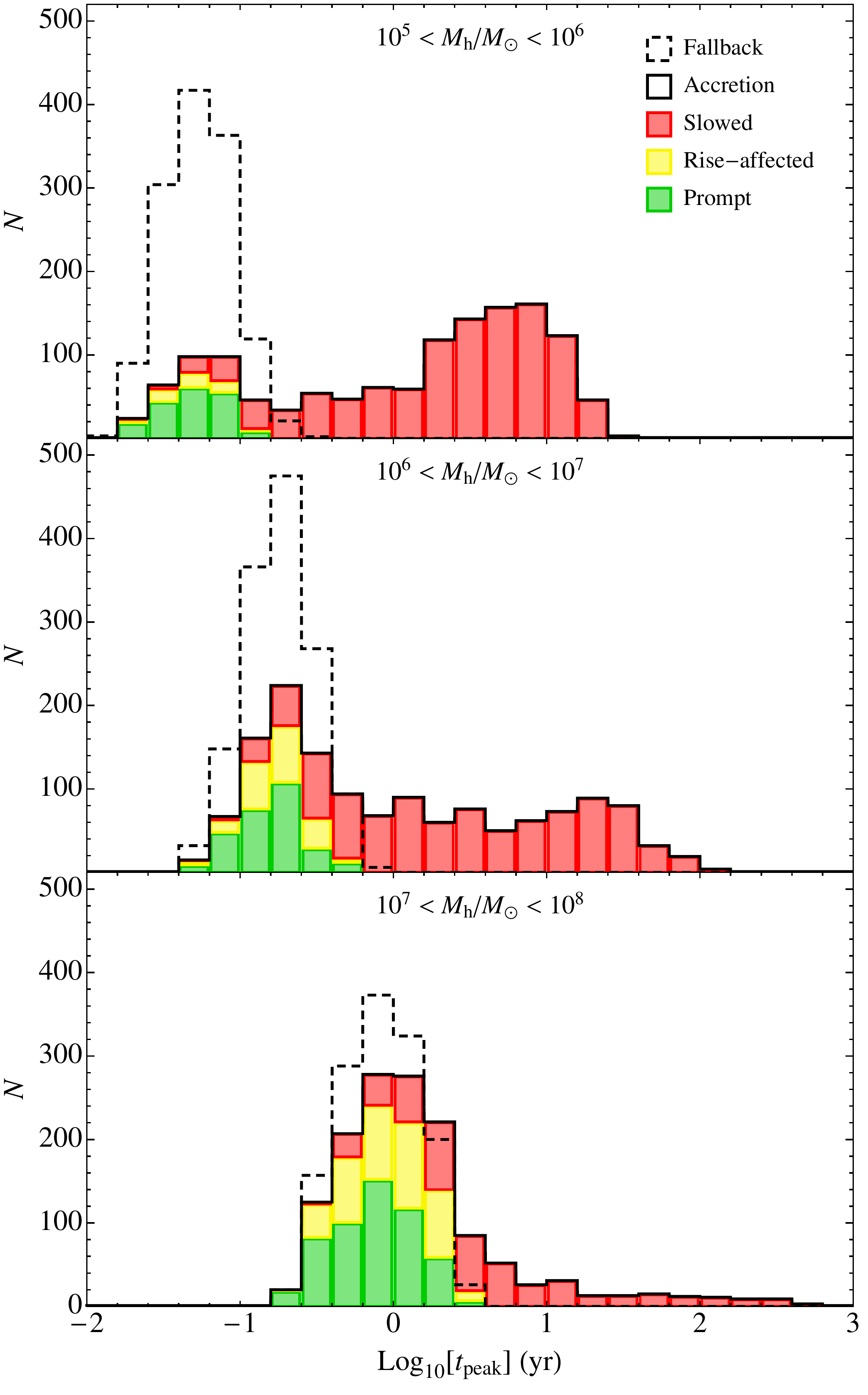}
\caption{Histograms showing the ratio of the peak to Eddington accretion rates \smash{$\dot{M}_{\rm peak}/\dot{M}_{\rm Edd}$} (left panel) and $t_{\rm peak}$ (right panel) for the initial material returning to pericenter \smash{$\dot{M}_{\rm fb}$} (dashed black line) and the accretion rate onto the black hole \smash{$\dot{M}_{\rm acc}$} (solid black line) for the three classes of TDEs defined in the text; color-coding matches that of Figures \ref{fig:delay} and \ref{fig:tdelay}. The three sub-panels show three different ranges of $M_{\rm h}$. In the left panel, the dotted vertical line shows where \smash{$\dot{M}_{\rm peak} = \dot{M}_{\rm Edd}$}.}
\label{fig:mpeak-tpeak}
\end{figure*}

Because no accretion takes place prior to circularization, this delay period is dark relative to the peak flare luminosity, with the only light likely coming from recombination of the stream as it expands, with luminosity $L \lesssim 10^{41}$~ergs~s$^{-1}$ \citep{Kasen:2010a}. This delay has implications for the time one expects to detect signatures of a disruption arising from the moment of maximum compression when the star first passes periapse, such as shock break-out \citep{Brassart:2008a,Guillochon:2009a} and gravitational wave emission \citep{Kobayashi:2004a,Rosswog:2009a,Stone:2013a}. For TDEs without a significant delay, these signatures would be expected to be detectable weeks to a few months prior to the peak of the accompanying accretion flare. Delayed events, on the other hand, can show these compression signatures years prior to the flare.

Figure \ref{fig:tdelay} also shows that slowed and prompt events co-exist for black holes of the same mass, indeed there are a small fraction of events that are prompt even for $M_{\rm h} = 10^{5} M_{\odot}$. Around low-mass black holes ($M_{\rm h} \lesssim 10^{6} M_{\odot}$) the tidal radius is significantly larger than $r_{\rm g}$ for main-sequence star disruptions. Therefore, prompt flares about these black holes require deeply plunging encounters, with $\beta \gtrsim 5$ being typical.

Because the Eddington limit $L_{\rm Edd} = 4\pi G M_{\rm h} m_{\rm p} c / \sigma_{\rm T}$ for a black hole scales with $M_{\rm h}$, TDEs about less-massive black holes tend to have peak fallback rates that are a few orders of magnitude greater than this limit, whereas more-massive black holes are usually sub-Eddington near peak. The slowing of flares around lower-mass black holes reduces their peak accretion rate relative to the fallback rate, and prolongs the flares such that they peak at a later time, whereas a large fraction of disruptions about higher-mass black holes are prompt, meaning that their accretion rates are not strongly affected by circularization. This has the net effect of reducing the peak accretion rate such that super-Eddington events are significantly more rare than if all TDEs were prompt.

In Figure \ref{fig:mpeak-tpeak} we show $\dot{M}_{\rm peak}/\dot{M}_{\rm Edd}$ and $t_{\rm peak}$, where we have presumed that the accretion efficiency $\mu = 1-\sqrt{1-2/3r_{\rm ISCO}}$ and $r_{\rm ISCO}$ is calculated as in \citet{Bardeen:1972a}, and we use the fitting relations of \citet{Guillochon:2013a} to determine $\dot{M}_{\rm peak}$ and $t_{\rm peak}$. By assuming that the peak accretion rate is reduced by a factor that is inversely proportional to $t_{\rm visc}/t_{\rm peak}$, we find 56\% of TDEs are super-Eddington once this slowdown effect is included for $10^{5}\,M_{\odot}\,<\,M_{\rm h}\,<\,10^{6}\,M_{\odot}$ versus 82\% if all TDEs were prompt, 34\% for $10^{6}\,M_{\odot}\,<\,M_{\rm h}\,<\,10^{7}\,M_{\odot}$ (versus 49\%), and 2.6\% for $10^{7}\,M_{\odot}\,<\,M_{\rm h}\,<\,10^{8}\,M_{\odot}$ (versus 4.0\%). Because more-massive black holes have higher Eddington luminosities, sub-Eddington events from higher-mass black holes will typically be more luminous than flares from lower-mass black holes. As the typical Eddington ratio only scales weakly with $M_{\rm h}$ for $M_{\rm h} < 10^{7}$ (as shown in the top two left panels of Figure \ref{fig:mpeak-tpeak}), and the Eddington accretion rate scales as $M_{\rm h}$, TDEs from black holes from $10^{5}$--$10^{7} M_{\odot}$ end up with similar peak accretion rates at peak, with a median value of $\dot{M}_{\rm peak} = 10^{-1.7} M_{\odot} {\rm yr}^{-1}$ over this mass range. Assuming a radiative efficiency of 10\%, this translates to a peak luminosity of $10^{44}$~ergs~s$^{-1}$, comparable to the inferred bolometric output of most observed TDEs. Because photons are largely trapped once the accretion rate exceeds Eddington \citep{Shakura:1973a}, the total disk luminosity is potentially limited to the Eddington luminosity \citep[though see][]{Jiang:2014a}, which would suggest that detections in those bands are further biased towards sub-Eddington events \citep{De-Colle:2012b}.

This slowdown also affects the timescale over which the flare evolves, which is shown in the right panel of Figure \ref{fig:mpeak-tpeak}. For lower-mass black holes, the slowdown is quite dramatic in most cases, causing some events to peak over timescales of several years. Given that most of the claimed optical/UV flares that were observed during maximum light peaked over a period of weeks or months \citep[e.g.][]{Gezari:2012a,Chornock:2014a,Arcavi:2014a,Holoien:2014a,Vinko:2015a}, and given that no tidal disruption candidates have been identified peaking on decade-long timescales, it is very possible that these slowed flares are mistaken for longer-term AGN activity. This may be the reason why tidal disruption surveys, which were originally predicted to detect dozens of flares per year \citep{Velzen:2011a}, have been discovering flares at a rate approximately one tenth that predicted from disruption rate estimates \citep{Wang:2004a,van-Velzen:2014a,Stone:2014a}. Moving to higher-mass black holes, the slowdowns affect a smaller fraction of the total, with the prompt fraction peaking on timescales of a few months to a year.

Because the prompt flares peak on shorter timescales, and because their bolometric light curves are likely to follow the original fallback rate $\dot{M}_{\rm fb}$ (thus preserving the $t^{-5/3}$ behavior), it is this sub-population of the TDEs that are likely to have been successfully identified as TDEs. The remainder, which are slowed significantly and are likely to have a time dependence that differs from the fallback rate predictions, may have been missed by current surveys. This argument also applies to the jetted TDEs \citep[e.g.][]{Bloom:2011a,Zauderer:2011a}, which require super-Eddington accretion rates to power the jet \citep{De-Colle:2012b} and rapid accretion timescales to explain the observed decay rate \citep{Tchekhovskoy:2014a,Kelley:2014a}.

\section{Summary}\label{sec:summary}
In this paper, we performed Monte Carlo realizations of tidal disruption streams to determine their structure prior to the onset of circularization. We have shown that the effects of GR are important when considering the evolution of the debris stream resulting from the tidal disruption of a star by a supermassive black hole.  When accounted for, we find that GR has the following effects:
\begin{itemize}
\item When GR effects are weak and the precession angles are small, the stream typically intersects itself far away from periapse. This results in a long viscous time that can potentially affect the accretion rate $\dot{M}_{\rm acc}$ and the time of peak accretion $t_{\rm peak}$ \citepalias{Guillochon:2014a}, a result that has been recently verified by numerical simulations \citep{Shiokawa:2015a,Bonnerot:2015a,Hayasaki:2015a}. The slowdown can result in $t_{\rm peak}$ increasing by a factor of $\sim$100, which also reduces $\dot{M}_{\rm acc}$ by a similar factor relative to the fallback rate $\dot{M}_{\rm fb}$. This mostly affects disruptions around black holes with masses less than a few $10^{6} M_{\odot}$.
\item For more-massive black holes, GR effects are significantly stronger, resulting in stream self-intersections that occur at large angles and closer to the star's original periapse. If the black hole spin is small ($a \lesssim 0.2$), the accretion rate onto the black hole typically follows the fallback rate, with no significant delay between the return of the most-bound material and accretion onto the black hole.
\item If a black hole is both massive and possesses at least a mild spin ($a \gtrsim 0.2$), the stream self-intersection occurs after the most-bound material has wound around periapse multiple times. Because very little energy is dissipated in the stream in the absence of stream self-intersections, the specific binding energy of the material to the black hole is unchanged until circularization begins, at which point the accretion rate will equal the fallback rate with a fixed time delay, $\dot{M}_{\rm acc}(t~+~t_{\rm delay})~=~\dot{M}_{\rm fb}(t)$.
\item Because tidal disruptions around less-massive black holes tend to be slowed down, and because more-massive black holes have higher Eddington luminosities, the typical Eddington ratio for a tidal disruption at peak luminosity is less than one, with most super-Eddington flares originating from black holes with masses $\lesssim 10^{7} M_{\odot}$.
\item TDEs that are significantly slowed by long viscous times likely peak on timescales of several years, with bolometric luminosities decaying at a rate more shallow than $t^{-5/3}$. If these slowed events were observed in the past, these differences might have prevented their identification as TDEs, as current surveys have focused on more-rapidly peaking transients. This may explain the relative dearth of events as compared to theoretical expectations.
\end{itemize}

The advantage of our approach is that we can generate a large ensemble of tidal disruption streams at little computational expense, but is heuristic in the sense that we made approximations to the precession terms using an orbit-averaged post-Newtonian formalism that introduces some error, as shown in Figure \ref{fig:compare}. A Monte Carlo approach that utilizes the exact solutions is computationally expensive for three reasons: First, we have found that the exact solutions take a few minutes to compute on a single CPU per realization, as compared to the near-instantaneous computation time (a fraction of a second) of the heuristic used here. Second, the problem of finding when and where a time-dependent parametric trajectory comes within a certain distance of itself is a three-dimensional root-finding problem, with the two positions on the curve and the time since disruption as free parameters. In our heuristic approach, time is discretized by the number of windings, so the self-intersection calculation is reduced to finding the solutions of $W_{1} (W_{1} - 1)$ two-dimensional root solves. Third, an even distribution of particles in energy does not result in a well-sampled elliptical spiral as the particles spend the majority of their time near apoapse. An accurate representation of the curve requires a careful selection of particle energies to represent the debris, with the optimal sample being different for every time $t$. The combination of these three issues mean that for every $t$ the full geodesics must be integrated from time zero using either a unique optimal sampling of particles (which takes minutes), or a large number of evenly spaced particles that guarantees the elliptical spiral will be well-sampled at all times. In either case, the total computation time is hours per realization, and is not suitable for generating an ensemble of thousands of systems. It is likely that alternative geodesic determinations that are faster to compute \citep[such as][]{Dexter:2009a} are required to make the exact Monte Carlo feasible.

We have asserted that the width of the debris stream grows via additional kinetic energy deposited each time the stream returns to periapse, but the evolution of the stream width as a function of time may be strongly influenced by cooling, recombination, shock heating, and magnetic fields. Cooling and recombination may cause the stream to fragment, resulting in the formation of discrete clumps within the stream \citep{Roos:1992a,Guillochon:2014b}, but will not affect the vertical spread in velocity as the stream will not be self-gravitating after its first return to periapse. However, heating of the stream through an oblique shock at periapse \citep{Matzner:2013a,Salbi:2014a} may increase this vertical spread, this would increase the typical stream widths and would result in streams that self-intersect after fewer windings than what is presented here. Lastly, the magnetic fields that permeate the star likely thread through the stellar debris and potentially become dynamically relevant once the gas pressure drops sufficiently, this additional pressure support may also increase the width of the stream.

\begin{figure}
\centering\includegraphics[width=0.8\linewidth,clip=true]{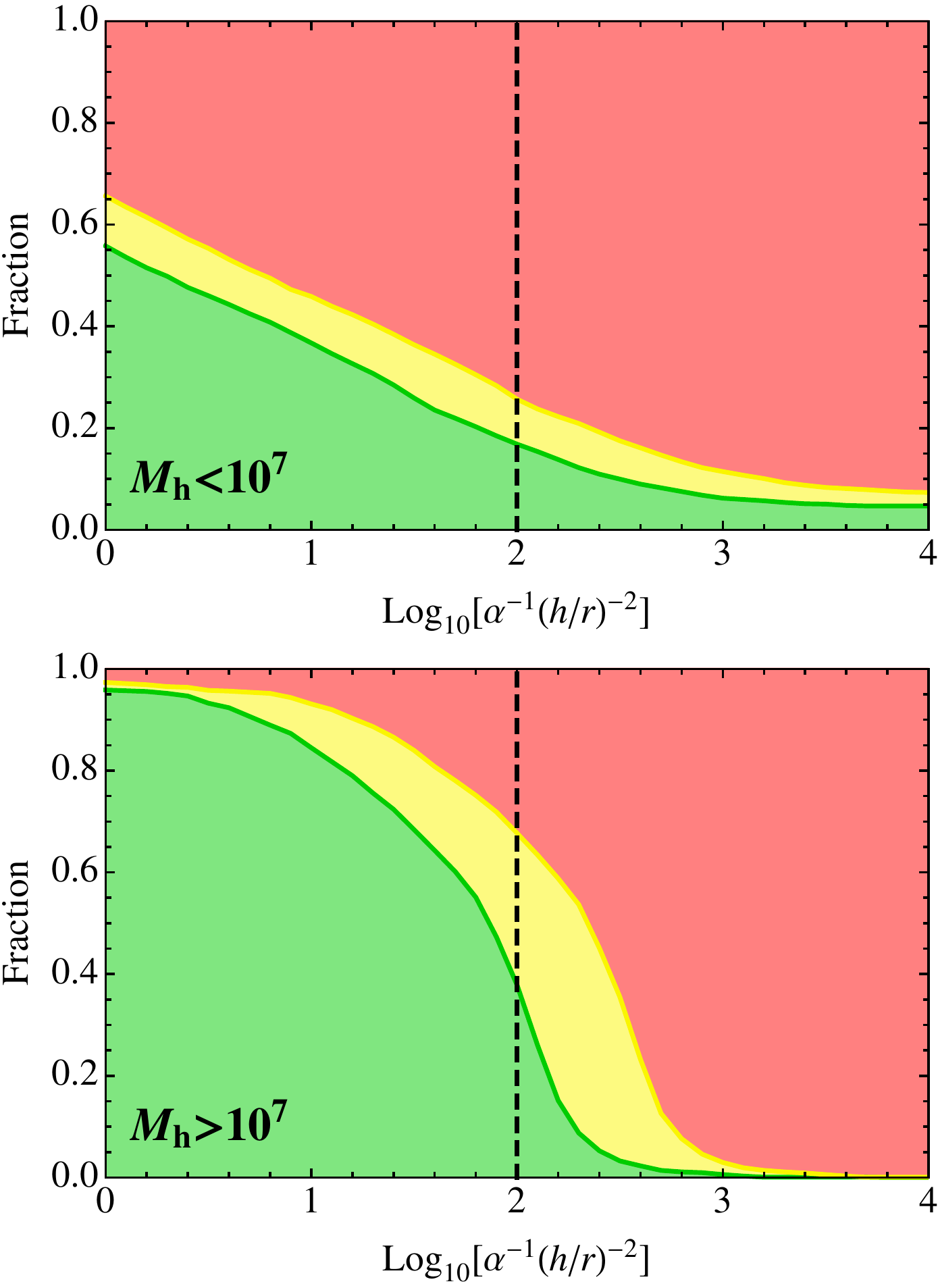}
\caption{Fraction of events that are prompt, rise-affected, and slowed as a function of $\alpha^{-1} (h/r)^{-2}$ below (top panel) and above (bottom panel) the break in delay times at $10^{7} M_{\odot}$. The vertical dashed line shows the fiducial value of $\alpha^{-1} (h/r)^{-2} = 100$ used to generate the previous plots and fitting relations presented in this paper.}
\label{fig:viscous}
\end{figure}

We have assumed that material circularizes when two streams intersect, but this may not be the case if the density of the two streams are wildly different from one another, as a denser portion of stream may uneventfully pass through a portion of significantly lower density. But because neighboring windings tend to be the ones we find intersecting (Figure~\ref{fig:windings}), the density ratio between the two colliding streams is expected to be similar. One improvement that could be made to our calculation would be to map ${\rm d}m/{\rm d}e$ to each ellipse to determine their relative densities.

Our results are also contingent upon the viscous evolution of the disk, and in this paper we assumed that $\alpha^{-1} (h/r)^{-2}~=~100$, a value typical of thick disks about steadily-accreting AGN. However, viscous times that are both shorter and longer than this fiducial value are possible. In thin accretion disks, $\alpha \sim 0.1$ and $(h/r)^{2} \sim 10^{-3}$ resulting in $\alpha^{-1} (h/r)^{-2} = 10^{4}$ \citep{King:2007a}, whereas ADAF flows can have $\alpha = 0.3$ \citep{Narayan:2008a} and $(h/r)^{2} \sim 0.25$ \citep{Yuan:2014a} resulting in $\alpha^{-1} (h/r)^{-2} = 10$. In Figure \ref{fig:viscous} we vary the value of $\alpha^{-1} (h/r)^{-2}$ and calculate the fraction of events that are prompt, rise-affected, and slowed. For disks that are more viscous than our fiducial assumption, we find the number of prompt flares increases with decreasing $\alpha^{-1} (h/r)^{-2}$, although this ratio is maximally 55\% (95\%) below (above) the break in delay time at $10^{7} M_{\odot}$ even for a highly viscous disk where $\alpha^{-1} (h/r)^{-2} = 1$. For less viscous disks, the prompt fraction drops to near zero for black holes above the delay time break, and accounts for only 5\% of flares below the break. While it is not clear if the value of $\alpha^{-1} (h/r)^{-2}$ depends on the fallback rate exceeding Eddington as it does for steadily-accreting AGN, the strong dependence of the prompt fraction on this value (especially for massive black holes) suggests that flares require $\alpha^{-1} (h/r)^{-2} \lesssim 100$ to be observable.

Despite these uncertainties, our work presented here demonstrates that black hole spin plays an important role in the circularization and eventual accretion of matter by supermassive black holes that have recently tidally disrupted a main-sequence star. These effects may also be important for other stellar disruptions in which $r_{\rm p}$ is comparable to $r_{\rm g}$, such as the disruption of a white dwarf by an intermediate mass black hole \citep{Rosswog:2009a,Krolik:2011a,Haas:2012a,MacLeod:2014b}, or the disruption of a giant star by a very massive ($\gtrsim 10^{8} M_{\odot}$) black hole \citep{MacLeod:2012a,MacLeod:2013a}. Future (magneto-)hydrodynamical simulations should aim to establish how the width of the stream evolves through the repeated periapse encounters made possible by the deflection induced by black hole spin, and the viscous evolution of the debris once it has circularized.

\bigskip
\acknowledgements
We thank J.~Dexter, G.~Farrar, A.~Gruzinov, Y.~Jiang, D.~Kasen, J.~Krolik, A.~Loeb, M.~MacLeod, M.~McCourt, M.~Rees, E.~Rossi, N.~Roth, J.~Steiner, N.~Stone, and E.~Tejeda for useful discussions, and the Aspen Center for Physics (NSF grant 1066293) for their hospitality. We also would like to thank the anonymous referee for their helpful suggestions. This work was supported by Einstein grant PF3-140108 (J.~G.), the Packard grant (E.~R.), and NASA ATP grant NNX14AH37G (E.~R.).

\bibliographystyle{apj}
\bibliography{/Users/james/Dropbox/library}

\end{document}